\documentclass[a4paper,11pt]{article}
\usepackage{pos,graphicx}

\def\simge{\mathrel{%
       \rlap{\raise 0.511ex \hbox{$>$}}{\lower 0.511ex \hbox{$\sim$}}}}
\def\simle{\mathrel{
       \rlap{\raise 0.511ex \hbox{$<$}}{\lower 0.511ex \hbox{$\sim$}}}}

\title{Latent heat and pressure gap at the first-order deconfining phase transition of SU(3) Yang-Mills theory using the small flow-time expansion method%
\\ \vspace*{-60mm}\hspace{4.9cm} \small{\texttt{UTHEP-762, UTCCS-P-141, J-PARC-TH-0253, KYUSHU-HET-228}} \vspace*{58mm}
}
 \ShortTitle{Latent heat and pressure gap of SU(3) YM using the SF$t$X method}

\author*[a]{Kazuyuki Kanaya}
\author[b]{Mizuki Shirogane}
\author[c]{Shinji Ejiri}
\author[d]{Ryo Iwami}
\author[e,f]{Masakiyo Kitazawa}
\author[g]{Hiroshi Suzuki}
\author[h]{Yusuke Taniguchi}
\author[i]{Takashi Umeda}

\affiliation[a]{Tomonaga Center for the History of the Universe, University of Tsukuba,\\ 
Tsukuba, Ibaraki 305-8571, Japan}

\affiliation[b]{Graduate School of Science and Technology, Niigata University, \\
Niigata 950-2181, Japan}

\affiliation[c]{Department of Physics, Niigata University, \\
Niigata 950-2181, Japan}

\affiliation[d]{Track Maintenance of Shinkansen, Rail Maintenance 1st Dept., East Japan Railway Co. Niigata Branch,\\ Niigata 950-0086, Japan}

\affiliation[e]{Department of Physics, Osaka University, \\ Toyonaka, Osaka 560-0043, Japan}

\affiliation[f]{J-PARC Branch, KEK Theory Center, Institute of Particle and Nuclear Studies, KEK,\\ 203-1, Shirakata, Tokai, Ibaraki, 319-1106, Japan}

\affiliation[g]{Department of Physics, Kyushu University,\\ 744 Motooka, Nishi-ku, Fukuoka 819-0395, Japan}

\affiliation[h]{Center for Computational Sciences, University of Tsukuba,\\ Tsukuba, Ibaraki 305-8577, Japan}

\affiliation[i]{Graduate School of Humanities and Social Sciences, Hiroshima University,\\ Higashihiroshima, Hiroshima 739-8524, Japan}

\emailAdd{kanaya@ccs.tsukuba.ac.jp}
\abstract{We study the latent heat and the pressure gap between the hot and cold phases at the first-order transition temperature $T=T_c$ of SU(3) Yang-Mills theory, using the small flow-time expansion (SF$t$X) method based on the gradient flow. 
We first examine alternative procedures in the SFtX method --- the order of the continuum and vanishing flow-time extrapolations. 
We confirm that the final results adopting the two orders, as well as other alternatives in which the perturbative order of the matching coefficients and the renormalization scale of the flow scheme are varied, are all consistent with each other.
We also confirm $\Delta p$ is consistent with zero, as expected from the dynamical balance of two phases at~$T_c$.
For the latent heat in the continuum limit, we find $\Delta \epsilon /T^4 = 1.117(40)$ for the spatial volume $L^3$ corresponding to the aspect ratio $N_s/N_t=T_cL=8$ and $1.349(38)$ for $N_s/N_t=6$. 
From hysteresis curves, we show that the entropy density in the hot phase is sensitive to the spatial volume, while that in the confined phase is insensitive. 
}

\FullConference{%
 The 38th International Symposium on Lattice Field Theory, LATTICE2021
  26th-30th July, 2021
  Zoom/Gather@Massachusetts Institute of Technology
}

\begin{document}
\maketitle

%%%%%%%%%%%%%%%%%%%%%%%%%%%%%%
\section{SF$t$X method based on the gradient flow}

\begin{figure}[b]
  \centering
%    \vspace*{-1mm}
    \includegraphics[width=0.95\textwidth]{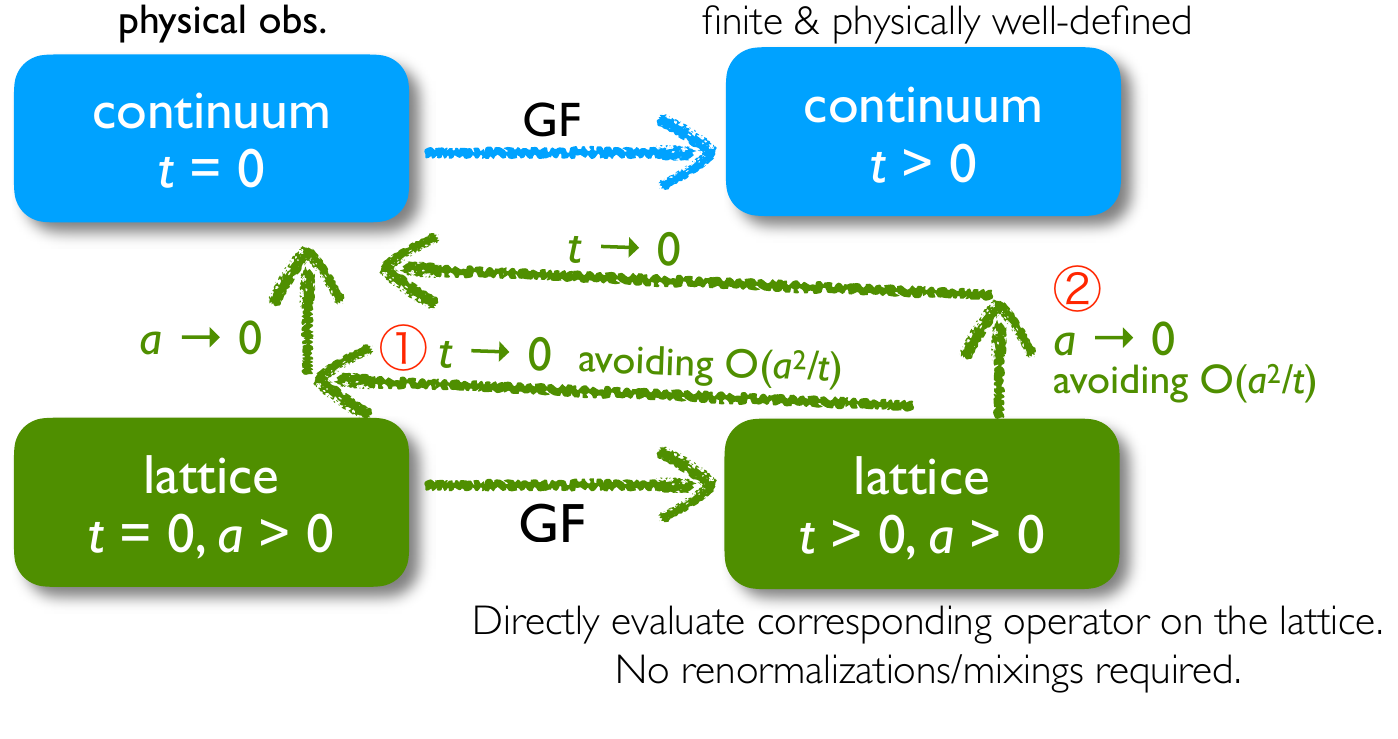}
    \vspace*{-3mm}
  \caption{
    Basic idea of the small flow-time expansion (SF$t$X) method~\cite{Suzuki:2013gza}.
  }
\label{fig:sftx}
\end{figure}

The deconfining phase transition of finite temperature SU(3) Yang-Mills theory provides us with a good testing ground for developing numerical techniques to investigate first order phase transitions in high-density and/or many-flavor QCD. 
In this report, we study it adopting the small flow time expansion (SF$t$X) method~\cite{Suzuki:2013gza,Makino:2014taa} based on the gradient flow\cite{Narayanan:2006rf,Luscher:2009eq,Luscher:2010iy,Luscher:2011bx}.
Using the fact that the fields at flow time $t>0$ are free from the ultraviolet divergences and short-distance singularities,
the SF$t$X method provides us with a general method to correctly calculate any renormalized observables on the lattice. 
The basic idea of the SF$t$X method is shown in Fig.~\ref{fig:sftx}: 
Because we can make any observable strictly finite by replacing the field variables in the observable with their flowed fields at $t>0$, we can calculate their non-perturbative values by simply evaluating their corresponding operator on the lattice. 
Here, unlike the cases of conventional lattice evaluation, we do not need to carry out numerical renormalization nor removal of contamination from unwanted operators due to lattice violation of relevant symmetry. 
Though the flowed observable is not the observable itself, we can get the latter by extrapolating the result of the flowed observable to the vanishing flow time limit $t\to0$.
The SF$t$X method has been successfully applied to evaluate
thermodynamic quantities in the Yang-Mills gauge theory~\cite{Asakawa:2013laa,Kitazawa:2016dsl,Hirakida:2018uoy,Iritani2019} and in QCD with $2+1$ flavors of dynamical quarks~\cite{WHOT2017b,WHOT2017,Lat2019-kanaya,Taniguchi:2020mgg}.

In this study, we adopt the SF$t$X method to calculate the latent heat $\Delta\epsilon$ and pressure gap $\Delta p$ at the first order phase transition of the SU(3) Yang-Mills theory~\cite{shirogane}.
Performing simulations at several lattice spacings and spatial volumes, we carry out the continuum extrapolation $a\to0$ and study the finite volume effect.

Another objective of this study is to test several alternative procedures for the SF$t$X method. 
In particular, we study the order of the double extrapolation $(a,t)\to0$ in the SF$t$X method: 
method~1 ($t\to0$ then $a\to0$) and method~2 ($a\to0$ then $t\to0$). See Fig.~\ref{fig:sftx}.
At $(a,t)\ne0$, we have lattice artifacts of ${\cal O}(a^2/t)$, which makes a naive $t\to0$ extrapolation difficult at $a\ne0$.
Though these artifacts are removed when we take $a\to0$ first, 
a reliable $a\to0$ extrapolation is in practice difficult at vary small $t$ where ${\cal O}(a^2/t)$ dominates the data. 
We thus have to carry out the double extrapolation avoiding regions where the ${\cal O}(a^2/t)$ artifacts are large. 
On the other hand, when the ${\cal O}(a^2/t)$ artifacts are avoided, the final results should be insensitive to the order of the two extrapolations~\cite{WHOT2017b}. 
The consistency of the methods 1 and 2 provides us with a good test of the numerical procedure to avoid the ${\cal O}(a^2/t)$ lattice artifacts. 

%%%%%%%%%%%%%%%%%%%%%%%%%%%%%%
\section{Setup}

Our flow equation is given by
\begin{equation}
   \partial_tB_\mu^a(t,x)=D_\nu G_{\nu\mu}^a(t,x) \equiv
\partial_\nu G_{\nu\mu}^a(t,x)+f^{abc}B_\nu^b(t,x)G_{\nu\mu}^c(t,x) 
\label{eq:(1.5)}
\end{equation}
with $B_{\mu}^a(0,x)=A_{\mu}^a(x)$, where $B_{\mu}^a(t,x)$ is the flowed gauge field and 
$
G_{\mu\nu}^a(t,x) \equiv \partial_\mu B_\nu^a(t,x)-\partial_\nu B_\mu^a(t,x)+f^{abc}B_\mu^b(t,x)B_\nu^c(t,x)
$
is the flowed field strength~\cite{Luscher:2009eq}.
The energy-momentum tensor (EMT) is then given by
\begin{eqnarray}
T_{\mu\nu}^R(x)
&=&\lim_{t\to0}\left\{ c_1(t) \,U_{\mu\nu}(t,x)
   +4c_2(t)\, \delta_{\mu\nu}
   \left[E(t,x)-\left\langle E(t,x)\right\rangle_0 \right]\right\},
\label{eq:EMT2}
\\
U_{\mu\nu}(t,x) &\equiv& G^a_{\mu\rho}(t,x)G^a_{\nu\rho}(t,x)
-\frac{1}{4}\delta_{\mu\nu}G^a_{\rho\sigma}(t,x)G^a_{\rho\sigma}(t,x),
\;\;
E(t,x) \equiv \frac{1}{4}G^a_{\mu\nu}(t,x)G^a_{\mu\nu}(t,x),
\nonumber
\end{eqnarray}
where $\left\langle \cdots \right\rangle_0$ is the zero temperature expectation value~\cite{Suzuki:2013gza}.
The matching coefficients $c_1(t)$ and $c_2(t)$ are expanded in perturbation theory as 
\begin{equation}
   c_1(t)=\frac{1}{g^2}\sum_{\ell=0}^\infty k_1^{(\ell)}(\mu,t)
   \left[\frac{g^2}{(4\pi)^2}\right]^\ell,
   \qquad
   c_2(t)=\frac{1}{g^2}\sum_{\ell=1}^\infty k_2^{(\ell)}(\mu,t)
   \left[\frac{g^2}{(4\pi)^2}\right]^\ell,
\label{eq:(1.6)}
\end{equation}
where $g=g(\mu)$ is the $\overline{\text{MS}}$ running coupling
and $\mu$ is the renormalization scale of the flow scheme.

The tree-level term is $k_1^{(0)}=1$, and $k_i^{(1)}$ and $k_i^{(2)}$ are one and two loop contributions given in Refs.~\cite{Suzuki:2013gza} and \cite{Harlander:2018zpi}, respectively.%
\footnote{
Note that our convention for $c_2(t)$ differs from that of Refs.~\cite{Harlander:2018zpi}. 
Our $c_2(t)$ corresponds to $c_2(t)+(1/4)c_1(t)$ in~Ref.~\cite{Harlander:2018zpi}.
}
In pure gauge Yang-Mills theories, $k_2^{(\ell+1)}$ can be deduced by $\ell$-loop coefficients using the trace anomaly~\cite{Suzuki:2013gza}. 
A concrete form for $k_2^{(3)}$ is given in Ref.~\cite{Iritani2019}. 
In this report, we show results adopting NNLO matching coefficients keeping terms up to $k_1^{(2)}$ for $c_1(t)$ and up to $k_2^{(3)}$ for $c_2(t)$. 
For the renormalization scale of the flow scheme, we adopt  $\mu_0(t) = 1 / \sqrt{2t e^{\gamma_{\text{E}}}}$ with $\gamma_{\text{E}}$ the Euler-Mascheroni constant proposed in Ref.~\cite{Harlander:2018zpi}.
This choice is reported to improve the signal of the SF$t$X method 
over the conventional choice $\mu_d(t) = 1/\sqrt{8t}$~\cite{Taniguchi:2020mgg}.

We study the energy density and the pressure obtained from the EMT as
\begin{equation}
\epsilon = -\left\langle T_{00}^R(x)\right\rangle, \qquad
p = \frac{1}{3} \sum_{i=1,2,3}\langle T_{ii}^R(x)\rangle.
\end{equation}
We note that the trace anomaly $\epsilon -3p$ is computed by the operator $E(t,x)$ 
with the matching coefficient $c_2(t)$, while the entropy density 
$\epsilon +p$ is computed by the operator $U_{\mu \nu}(t,x)$ with $c_1(t)$.
We thus calculate these conventional combinations too.

We perform simulations with the standard Wilson action at several $\beta$'s around the transition point $\beta_c$ 
on lattices with $N_t=8$, $12$ and $16$ with spatial lattice sizes $N_s$ corresponding to the aspect ratio $N_s/N_t = 6$ and 8. For $N_t=12$ we also simulate $N_s=48$ and 64 lattices.
Because $T$ is adjusted to $T_c$ in our study, the lattice spacing $a = 1/(N_t T_c)$ is fixed by $N_t$, 
and the spatial volume in physical units, $L^3 = (N_s a)^3 = N_s^3/(N_t T_c)^3$, is fixed by the aspect ratio $N_s/N_t=T_cL$.

We define the transition point $\beta_c$ as the peak position of the Polyakov loop susceptibility 
$\chi_{\Omega}=N_s^3 (\langle \Omega^2 \rangle -\langle \Omega \rangle ^2)$.
Near the first order transition point, we separate the configurations into the hot (deconfined) and cold (confined)  phases using the spatially averaged Polyakov loop $\Omega$.
See \cite{shirogane} for details.
After the phase separation, we carry out the gradient flow on each of the configurations to measure flowed operators. 
In each of the hot and cold phases, we then combine the expectation values of flowed operators at different simulation points by the multipoint reweighting method, to obtain the values just at the transition point~$\beta=\beta_c$ (\textit{i.e.} $T=T_c$),
and calculate the gaps $\Delta\epsilon/T^4$, $\Delta (\epsilon -3p)/T^4$, $\Delta (\epsilon +p)/T^4$ and $\Delta p/T^4$ between the hot and cold phases at the first-order transition temperature $T_c$.
When $\Delta p=0$ as expected from the dynamical balance between the two phases at $T_c$, the latent heats
$\Delta\epsilon/T^4$, $\Delta (\epsilon -3p)/T^4$ and $\Delta (\epsilon +p)/T^4$ should coincide with each other. 

%%%%%%%%%%%%%%%%%%%%%%%%%%%%%%
\section{Results}

\begin{figure}[t]
  \centering
    \vspace*{-1mm}
    \includegraphics[width=0.495\textwidth]{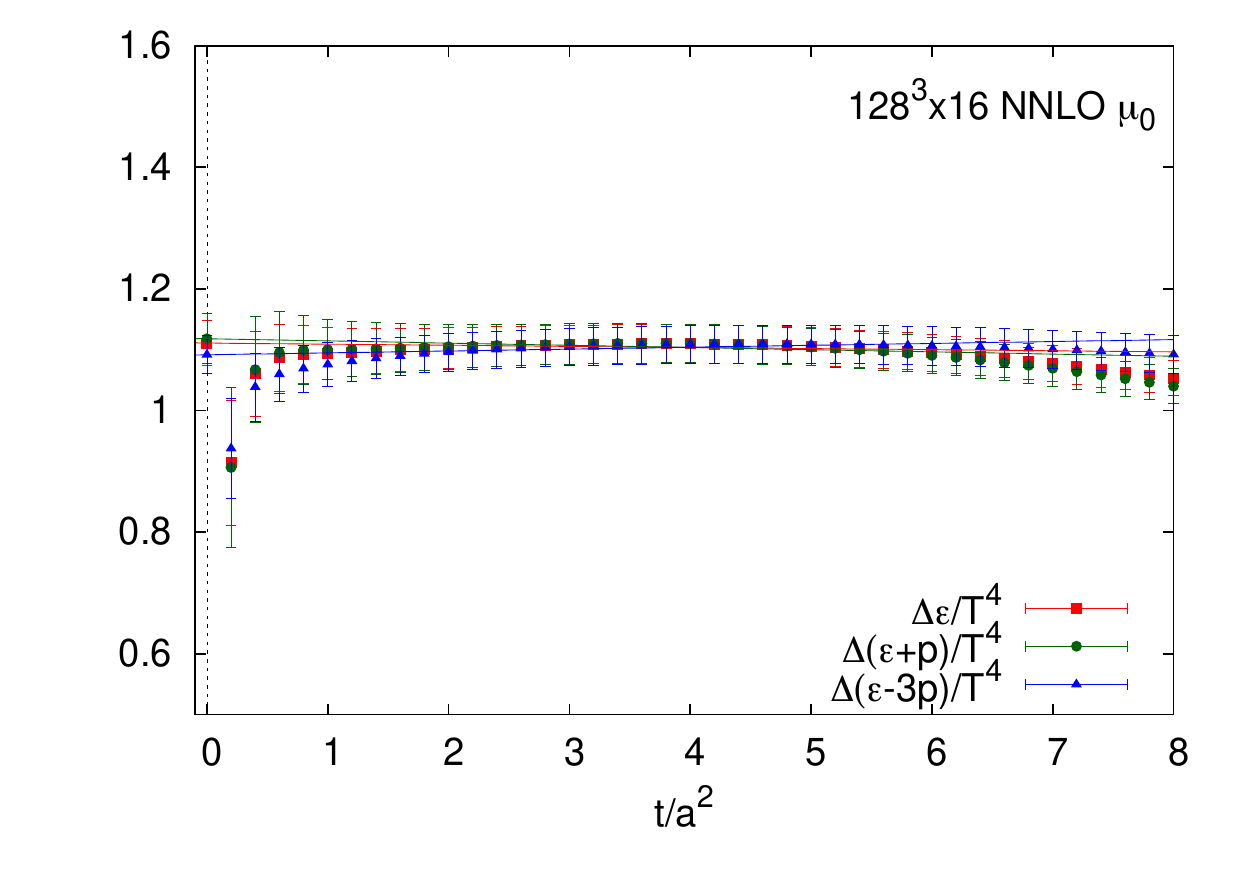}
%    \hspace*{0.5mm}
    \includegraphics[width=0.495\textwidth]{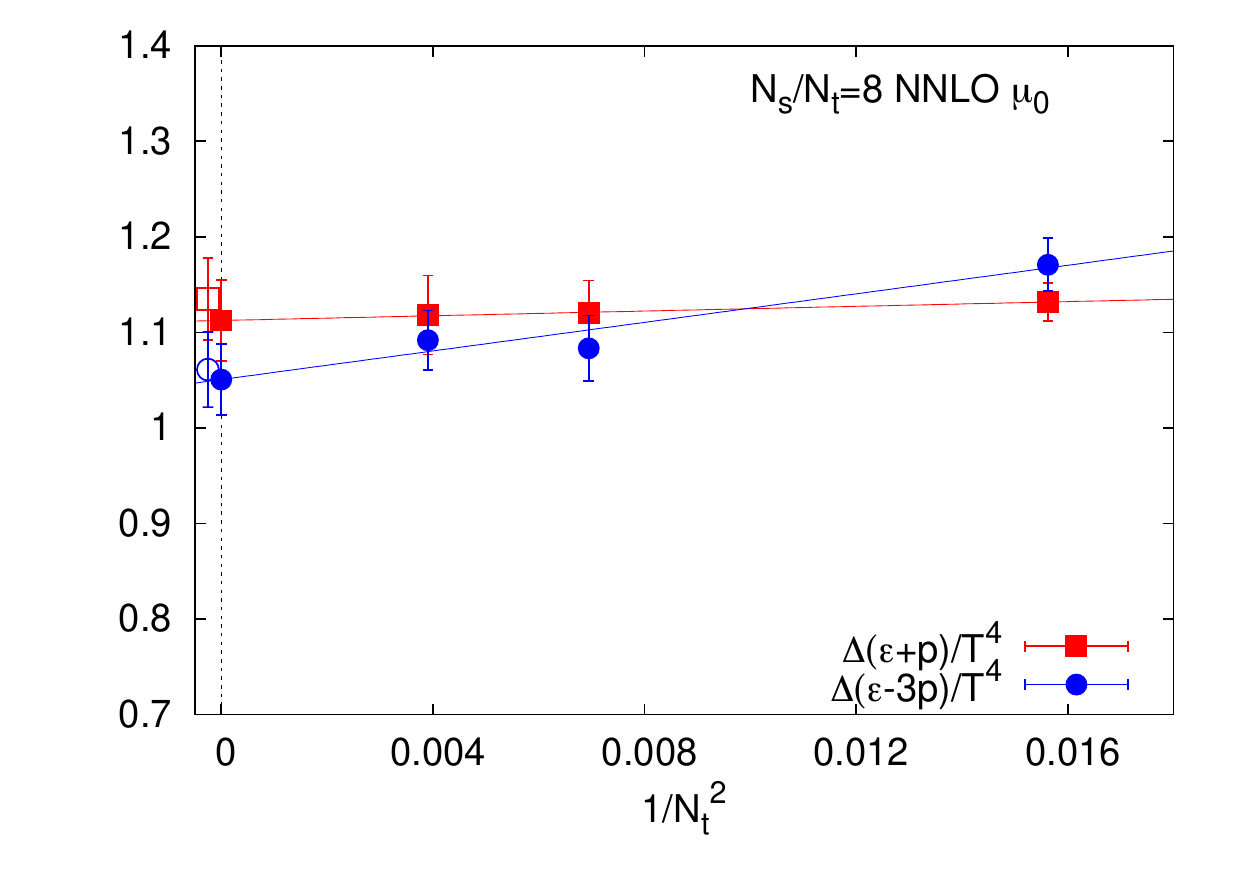}
    \vspace*{-4mm}
  \caption{
    Double extrapolation of the latent heat by the method~1. 
    \textbf{Left:} $t\to0$ extrapolation of $\Delta\epsilon/T^4$, $\Delta(\epsilon+p)/T^4$ and $\Delta(\epsilon-3p)/T^4$ obtained on the $128^3\times16$ lattice.
    \textbf{Right:} $a\to0$ extrapolation of the results at $t=0$ on $N_s/N_t=8$ lattices.~\cite{shirogane}
  }
\label{fig:m1}
\end{figure}

Figure~\ref{fig:m1} shows our determination of the latent heat by the method~1.
The left panel is the $t\to0$ extrapolation of $\Delta\epsilon/T^4$, $\Delta(\epsilon+p)/T^4$, and $\Delta(\epsilon-3p)/T^4$ at fixed $a$, obtained on the $128^3\times16$ lattice as an example.
The rapid change of data at $t/a^2 \simle 0.5$ is due to the ${\cal O}(a^2/t)$ lattice artifacts.
Note that, when we remove the small and large $t$ regions where ${\cal O}(a^2/t)$ and ${\cal O}(t^2)$ effects are appreciable, the results of $\Delta\epsilon/T^4$, $\Delta(\epsilon+p)/T^4$ and $\Delta(\epsilon-3p)/T^4$ are all consistent with each other, meaning that $\Delta p$ is consistent with zero already at $(a,t)\ne0$. 
Selecting a linear window in which the data is well linear, we perform linear extrapolations to $t=0$, shown by straight lines and the symbols at $t=0$ in the left panel.
See Ref.~\cite{shirogane} for our criterion to choose an optimum window. 
We confirm that the results are consistent within statistical errors under a variation of fitting windows.
We then perform $a\to0$ extrapolation at each fixed spatial volume, as shown in the right panel of Fig.~\ref{fig:m1} for the case of the spatial volume corresponding to $N_s/N_t=8$.
The horizontal axis is $1/N_t^2 = (aT_c)^2$.
The linear lines with the symbols at $a=0$ are the results of $a\to0$ extrapolation using data at three lattice spacings corresponding to $N_t=8$, $12$ and $16$.

\begin{figure}[t]
  \centering
    \vspace*{-25mm}
    \includegraphics[width=0.52\textwidth]{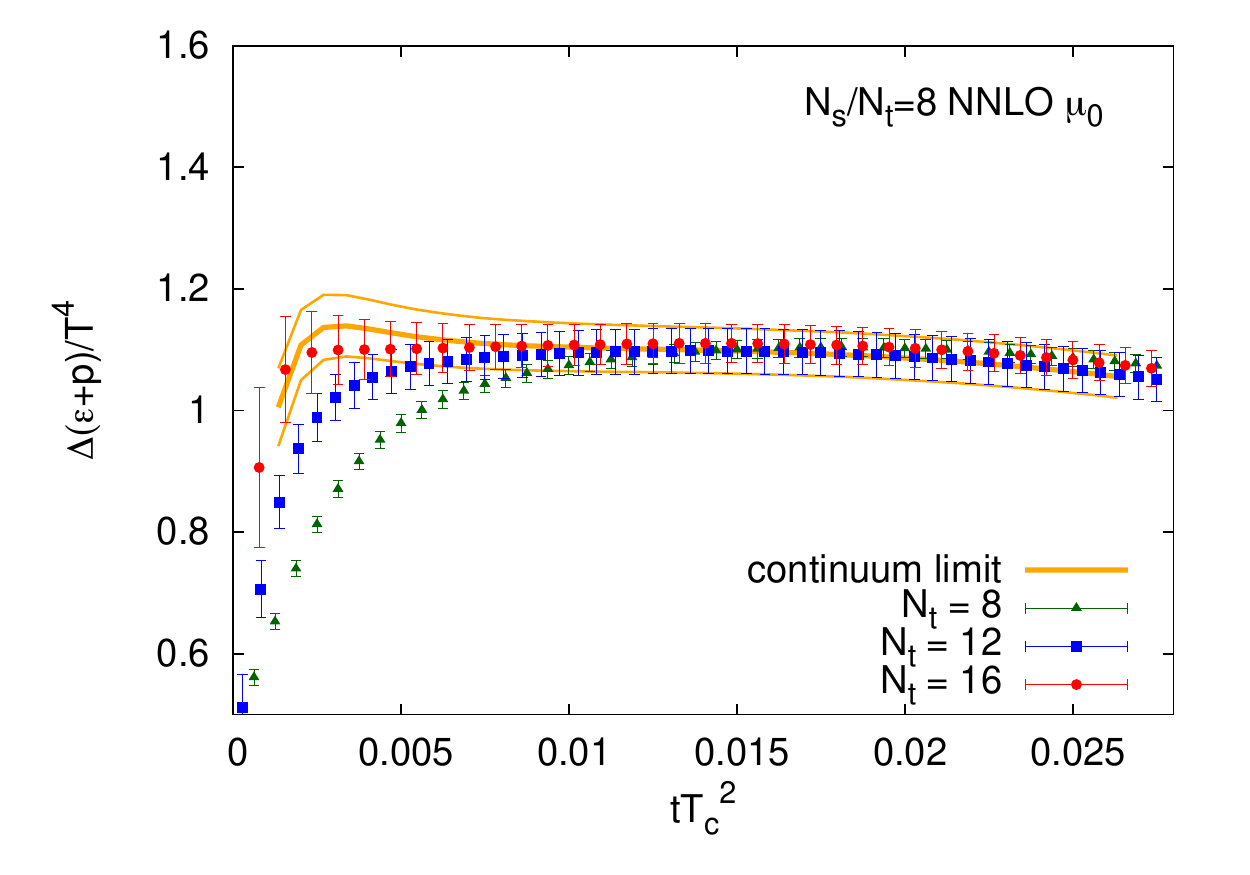}
%    \hspace*{-0.2mm}
    \includegraphics[width=0.47\textwidth]{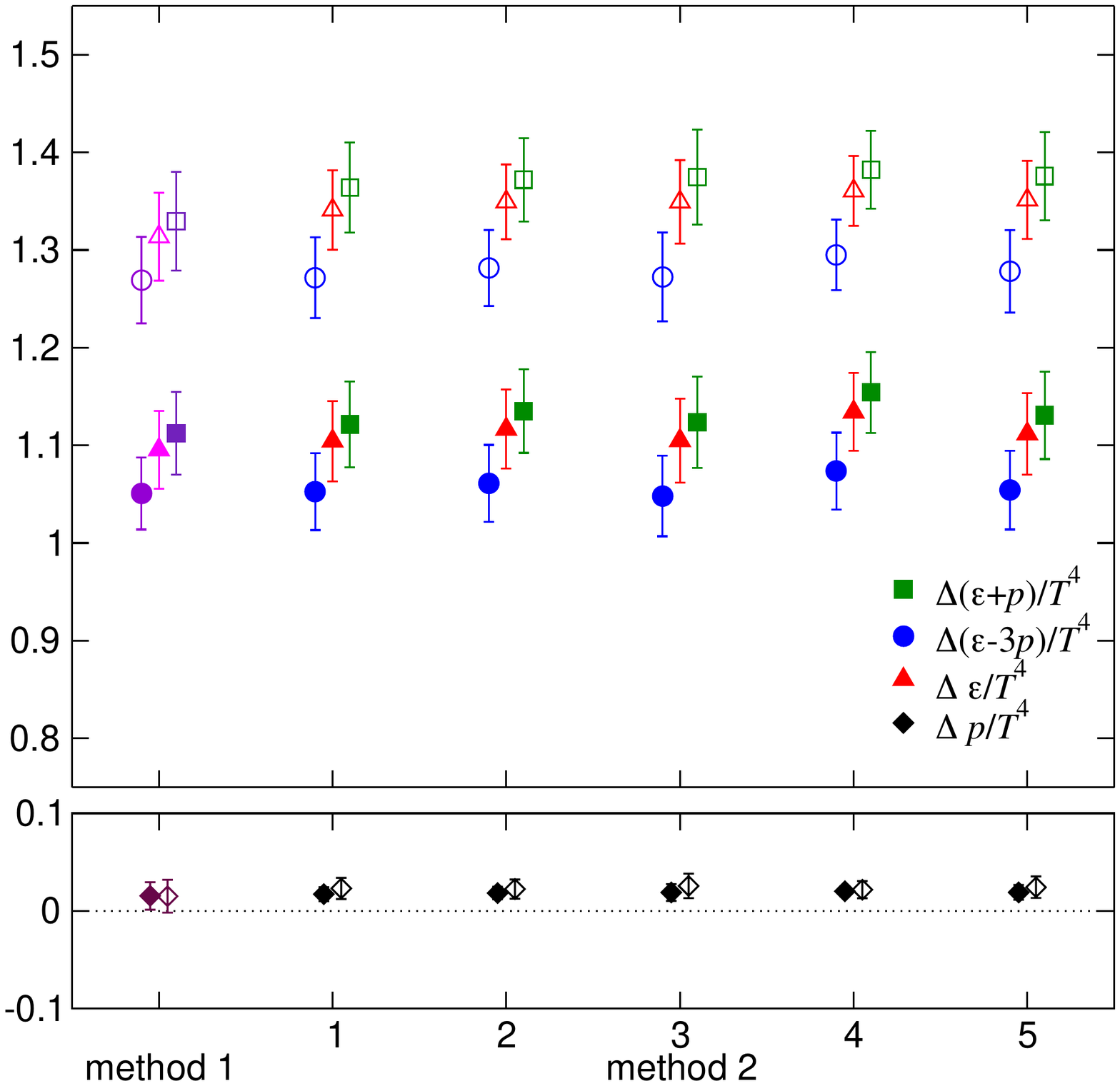}
    \vspace*{-3mm}
  \caption{
    Double extrapolation of the latent heat and the pressure gap by the method~2. 
    \textbf{Left:} $a\to0$ extrapolation of $\Delta(\epsilon+p)/T^4$ at each $t$ in physical units obtained on lattices with $N_s/N_t=8$.
    Results of the $a\to0$ extrapolation are shown by the thick orange curve,
    with which we perform $t\to0$ extrapolation to get the physical value of $\Delta(\epsilon+p)$.
    \textbf{Right:} Final results of $t\to0$ extrapolations using the results in the continuum limit, with various fit ranges denoted by the numbers 1--5 on the horizontal axis. 
    The filled and open symbols are for $N_s / N_t =8$ and $6$, respectively.  
    Results of the method~1 are also shown at the left end.
    See Ref.~\cite{shirogane} for details.
  }
\label{fig:m2}
\end{figure}

Figure~\ref{fig:m2} shows our results of latent heat and pressure gap by the method~2.
We first make $a\to0$ extrapolation at each $t$ in physical units using data at three $a$ corresponding to $N_t=8$, $12$ and $16$, fixing $N_s/N_t$ to take the finite size effect into account. 
The result of the linear $1/N_t^2$ extrapolation for $\Delta(\epsilon+p)/T^4$ at $N_s/N_t=8$ is shown by the thick orange curve in the left panel of Fig.~\ref{fig:m2}, with thin orange curves for the statistical error.
We then perform $t\to0$ extrapolation of the orange curve, selecting several candidates of the linear window. 
See Ref.~\cite{shirogane} for details.
The right panel of Fig.~\ref{fig:m2} show the fit range dependence of the results of $t\to0$ extrapolation. 
We also show the final results of method~1 at the left end of the plot. 
We find that the results adopting different fit ranges as well as those obtained by the method~1 are vary consistent with each other.

The results of $\Delta p$ by direct calculation from EMT are given at the bottom of the right panel of Fig.~\ref{fig:m2}. 
We find that the values of $\Delta p$ are only about 1\% of the latent heat and the results of the methods~1 and 2 are consistent with each other.
Due to correlation between $\Delta (\epsilon +p) /T^4$ and $\Delta (\epsilon -3p) /T^4$, the jackknife statistical errors for $\Delta p$ turned out to be quite small in comparison to the errors of the latent heat.
With the small errors, the mean values of $\Delta p$ deviate from zero by about 2-3 $\sigma$ statistical errors.
We find that these nonvanishing values in the $t\to0$ limit originate solely from the data at $N_t=8$ used in the continuum extrapolation:  
We find that the values of $\Delta p$ deviate from zero at $N_t=8$, while those at $N_t=12$ and 16 are consistent with zero in the linear windows for the $t\to0$ extrapolation. 
When we remove the data on the coarsest lattice $N_t=8$, we obtain $\Delta p$ consistent with zero. 
Thus, taking account of the systematic error due to the continuum extrapolation which is larger than the statistical error for the case of $\Delta p$, we conclude that the pressure gap $\Delta p$ is consistent with zero.

For the latent heat, adopting the results of the method~2 as central values, we obtain $\Delta \epsilon /T^4 = 1.117(40)$ for the spatial volume corresponding to $N_s/N_t=8$, 
and $1.349(38)$ for $N_s/N_t=6$.
Systematic errors estimated from the differences between the methods~1 and 2 as well as among different fit ranges are smaller than the statistical errors quoted here.

%%%%%%%%%%%%%%%%%%%%%%%%%%%%%%
\section{Hysteresis of entropy density around $T_c$}

In the previous section, we find that the latent heat is clearly dependent on the spatial volume of the system at $N_s/N_t=6$ and $8$. 
At a first-order phase transition point, however, because the correlation length does not diverge, the spatial volume dependence should be mild when the volume is sufficiently large.
To find the origin of the spatial volume dependence in our latent heat, we study the entropy density $(\epsilon + p) /T^4$ separately in the hot and cold phases at temperatures around $T_c$ using the multipoint reweighting method. 

\begin{figure}[t]
  \centering
    \vspace*{-2mm}
    \includegraphics[width=0.7\textwidth]{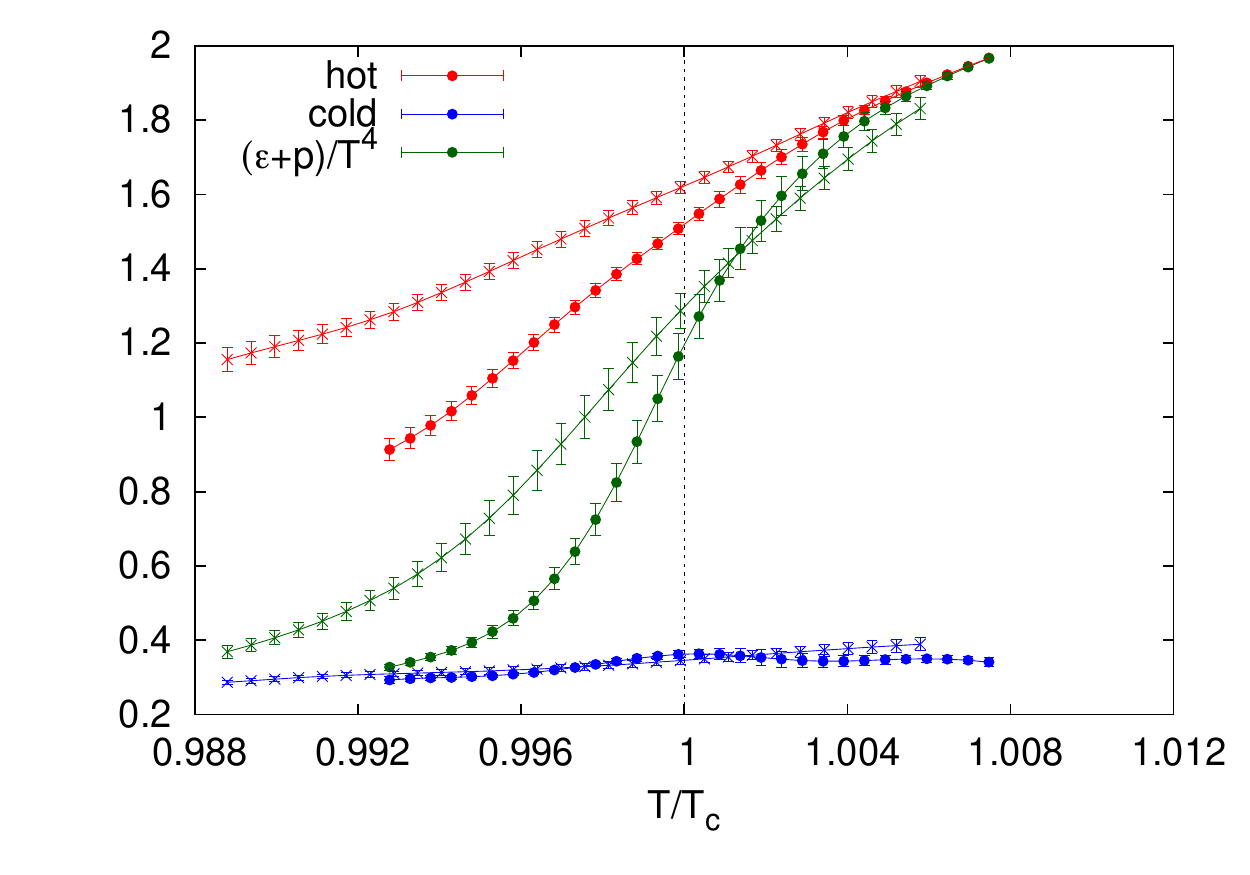}
%    \hspace*{2mm}
%    \includegraphics[width=0.45\textwidth]{figs/flow0lim_aspectlatio8_NL_u0.pdf}
    \vspace*{-2mm}
  \caption{
    $(\epsilon+p)/T^4$ by the SF$t$X method calculated using configurations in the hot phase (red), the cold phase (blue) and all configurations without the phase separation (green), obtained on the $48^3 \times 8$ (cross) and $64^3 \times 8$ (circle) lattices.~\cite{shirogane}
  }
\label{fig:hisflow}
\end{figure}

The results of $(\epsilon + p) /T^4$ obtained on $48^3 \times 8$ and $64^3 \times 8$ lattices are plotted by the cross and circle symbols in Fig.~\ref{fig:hisflow}. 
Results at $t/a^2=1.4$ are shown --- the results in the $t\to0$ limit are about the same with slightly larger errors. 
The red symbols are for the results in the hot phase and the blue symbols in the cold phase, while the green symbols show the results without the phase separation.
The horizontal axis is the temperature $T=1/(N_t a)$ normalized by $T_c$, 
where the relation between $a$ and $\beta$ is determined by the critical point $\beta_c$ as function of $N_t$~\cite{shirogane16}. 

We find that the spatial volume dependence appears only in the (metastable) hot phase around $T_c$, while no apparent volume dependence is visible in the cold phase.
We thus conclude that the spatial volume dependence of the latent heat is due to that in the contribution of the hot phase.
From this figure, we also find that latent heat is sensitive to the value of the critical point $\beta_c$.
A careful determination is required for $\beta_c$.

Note that a clear identification of the spatial volume effect is enabled by the small errors by the SF$t$X method. 
In Ref.~\cite{shirogane}, we show a corresponding plot by the derivative method using the same configurations, in which the errors are much larger.
The small errors for the energy density and pressure by the SF$t$X method are in part due to the simpleness of the measurement procedure for the energy density and pressure --- in contrast to the cases of conventional integral or derivative methods, information of nonperturbative beta functions or Karsch coefficients are not needed in the SF$t$X method, 
and also due to the smearing nature of the gradient flow which naturally suppresses statistical fluctuations.

%%%%%%%%%%%%%%%%%%%%%%%%%%%%%%
\section{Conclusions}

We carried out a series of systematic simulations of SU(3) Yang-Mills theory including those at three lattice spacings and two spatial volumes, around the transition temperature $T_c$, and calculated the energy-momentum tensor applying the small flow-time expansion (SF$t$X) method based on the gradient flow. 
We fine-tuned the temperature to $T_c$ by the multipoint reweighting method, 
and calculated the energy density and the pressure from the energy-momentum tensor separately in metastable hot (deconfined) and cold (confined) phases just at $T_c$, to calculate the latent heat and the pressure gap.

Using the systematic data at various lattice spacings, we first examined the two alternatives for the double extrapolation $(a,t)\to(0,0)$ in the SF$t$X method --- method~1 (first $t\to0$ and then $a\to0$) and method~2 (first $a\to0$ and then $t\to0$). 
When the ${\cal O}(a^2/t)$ lattice artifacts are correctly avoided, the final results of the methods~1 and 2 should agree with each other. 
We found that the results of the latent heat and the pressure gap adopting the methods~1 and 2 agree well with each other.
In Ref.~\cite{shirogane}, we have also tested the influence of the truncation of the perturbative series for the matching coefficients by repeating the calculations with the NLO matching coefficients, 
and also that due to the choice of the renormalization scale $\mu$ by repeating the analyses with the conventional choice $\mu_d = 1/\sqrt{8t}$.
We confirmed that the final results with these alternative procedures are all consistent with each other, 
while the choice $\mu_0$ improves linear windows and the use of NNLO matching coefficients improves the signal of the latent heat in the sense that the pressure gap is more clearly suppressed~\cite{shirogane}.
These results ensure our numerical procedures for the SF$t$X method.

The final results for the pressure gap $\Delta p$ between the hot and cold phases turned out to be consistent with zero, as expected from the dynamical balance of two phases at~$T_c$. 
For the latent heat, we obtained $\Delta \epsilon /T^4 = 1.117 \pm 0.040$ for the spatial volume $L^3 = (N_s/N_t)^3 / T_c^3$ corresponding to the aspect ratio $N_s/N_t=8$, and $1.349 \pm 0.038$ for $N_s/N_t=6$.
From a study of hysteresis curves around $T_c$, we found that this spatial volume dependence is caused by that of the observables in the metastable hot phase.
The clear volume dependence of the latent heat calls for a study on larger spatial volumes with high statistics.
Importance of large spatial volume was emphasized also in a recent study of finite size scaling in QCD with heavy quarks~\cite{kiyohara}.

%%%%%%%%%%%%%%%%%%%%%%%%%%%%%%
\section*{Acknowledgments}
%\vspace{5mm}

We thank other members of the WHOT-QCD Collaboration for valuable discussions.
This work was in part supported by JSPS KAKENHI Grant Numbers JP21K03550, JP20H01903,
JP19H05146, JP19H05598, JP19K03819, JP18K03607, JP17K05442 and JP16H03982, 
and the Uchida Energy Science Promotion Foundation.
This research used computational resources of COMA, Oakforest-PACS, and Cygnus provided by the Interdisciplinary Computational Science Program of Center for Computational Sciences, University of Tsukuba,
K and other computers of JHPCN through the HPCI System Research Projects (Project ID:hp17208, hp190028, hp190036, hp200089) and JHPCN projects (jh190003, jh190063, jh200049), OCTOPUS at Cybermedia Center, Osaka University, ITO at Research Institute for Information Technology, Kyushu University, and Grand Chariot at Information Initiative Center, Hokkaido University, 
SR16000 and BG/Q by the Large Scale Simulation Program of High Energy Accelerator Research Organization (KEK) (Nos.\ 14/15-23, 15/16-T06, 15/16-T-07, 15/16-25, 16/17-05). 
%The authors also thank the Yukawa Institute for Theoretical Physics at Kyoto University for the workshop YITP-W-19-09.

%%%%%%%%%%%%%%%%%%%%%%%%%%%%%%

\end{document}